\documentclass[aps,amssymb,prl,twocolumn]{revtex4}
\usepackage[dvips]{graphicx}
\usepackage{latexsym}
\usepackage{pstricks,pst-node,pst-text,pst-3d}
\usepackage{pslatex}
\usepackage{epsfig}

\begin{document}
\title{Triangular lattice neurons may implement an advanced numeral system to precisely encode rat position over large ranges}
\author{Yoram Burak$^{1}$, Ted Brookings$^{2}$, Ila Fiete$^{1,*}$}
\affiliation{Kavli Institute for Theoretical Physics$^{1}$ and Department of Physics$^{2}$, University of California, Santa Barbara 93106\\ $^*$prasad@kitp.ucsb.edu}


\begin{abstract} We argue by observation of the neural data that neurons in area dMEC of rats, which fire whenever
the rat is on any vertex of a regular triangular lattice that tiles 2-d space,
may be using an advanced \emph{numeral system} to reversibly encode rat position. 
We interpret measured dMEC properties within the framework of a residue number system (RNS), and describe how RNS encoding -- which breaks the non-periodic variable of rat position into a set of narrowly distributed periodic variables -- allows a small set of cells to compactly represent and efficiently update rat position with high resolution over a large range. We show that the uniquely useful properties of RNS encoding still hold when the encoded and encoding quantities are relaxed to be real numbers with built-in uncertainties, and provide a numerical and functional estimate of the range and resolution of rat positions that can be uniquely encoded in dMEC. The use of a compact, `arithmetic-friendly' numeral system to encode a metric variable, as we propose is happening in dMEC, is qualitatively different from all previously identified examples of coding in the brain. We discuss the numerous neurobiological implications and predictions of our hypothesis. 

\noindent
\end{abstract}
\maketitle

Recent experiments reveal that rats, notoriously good navigators, encode
information about their position in 2-dimensional spaces in a remarkable way
\cite{GridCells1, GridCells2}. Each neuron in the dorsolateral band of the
mediolateral entorhinal cortex (dMEC) fires when the rat is on any vertex
of an imagined regular triangular lattice, tiling the plane. The firing pattern
is independent of the enclosure size and shape, and updates correctly as the rat
moves around, even in complete darkness. These observations, together with
lesion studies, hint that dMEC may play a central role in rodent path
integration \cite{Pathinteg1, Pathinteg2, GridCells2,Burak06}.

Nearby dMEC neurons share the same lattice period ($\lambda$) and orientation,
and differ only in their relative spatial phases \cite{GridCells2}. Thus at any
instant, the active subset of neurons only specifies current rat position as a
phase: upto, or \emph{modulo}, all possible periodic displacements of the lattice.  
Moving ventrally along the length of dMEC, the neural lattice period increases
monotonically, but the measured range is narrow: from approximately 30 cm to 70
cm \cite{GridCells2}.  In summary, we note that dMEC neurons decompose the 2-dimensional 
vector of rat position into a set of phases, modulo a set of lattice periods, along two 
independent lattice directions. 

While phases within the smallest lattice -- about the size of the rat -- might
arguably vary at the appropriate scale for fine position discrimination during
navigation, the largest lattice period ($<$2 m, by extrapolation along dMEC
length) puzzlingly appears to fall far short of the range over which a rat might 
know its position.

Can the phases of such a narrow range of lattices even theoretically contain enough
information to unambiguously represent position over the
behaviorally relevant range of distances covered by rats? And even if so, what
is the advantage of decomposing rat position -- a quantity that can in theory be
represented simply by a pair of numbers -- in this distributed, seemingly bizarre way?

Based on the observation that dMEC stores only modulo information about rat position, 
we suggest that dMEC may be encoding and enabling the reconstruction of 
the two-coordinate rat position vector according to a generalized version of a residue number 
system (RNS) \cite{RNSreview}, a scheme with uniquely useful properties 
in the neurobiological context. 
In an RNS, a number $x$ (e.g., position in 1-d) is represented by a list ($x_i = x \ {\rm
mod} \ \lambda_i$) of its \emph{residues}, or remainders after division by a set
of fixed numbers ($\lambda_1, \lambda_2, ..., \lambda_N$), relatively prime to
each other, called \emph{moduli}. For example, if the moduli are (13, 15, 16,
17, 19), the number 1000000 is represented by the residues (1, 10, 0, 9, 11). The
Chinese Remainder Theorem (CRT) guarantees that any number smaller than the
product of all the moduli is uniquely specified by (and therefore can be
reconstructed from) its residues. In the example above, any number up to 1007760
has a unique representation. 
An RNS has striking parallels to dMEC encoding:
According to our interpretation, the dMEC lattice periods are the RNS moduli 
(which need not be co-prime integers, as we discuss later); the lattice phases, 
specified by the active set of neurons at any given rat position, are the residues. 


In an RNS, large numbers (e.g. position in a large space) are represented 
combinatorially and thus compactly; this property is shared with base numeral systems like
decimal or binary. A small set of registers ($N$) can represent a very large
range of numbers ($[0,10^N-1]$: decimal, $[0,2^N-1]$: binary, or
$[0,~\lambda^N-1]$: RNS, where $\lambda$ is the approximate size of the moduli,
and $N$ is the number of moduli; in the example above, $\lambda \approx 16$, $N=5$). 
By contrast, in a sparse encoding scheme where each element represents one
possible number (e.g. place-cell representation of position in hippocampus), 
$N$ elements only cover the range $[0,N-1]$. But besides its combinatorial
capacity, two unique features of an RNS make it especially useful, compared to
base numeral and other systems, for position encoding in dMEC.

First, the set of registers in an RNS that support its vast representational
capacity can be very closely spaced, spanning a narrow range: as illustrated
above, all moduli may be nearly equal. By contrast, successive registers in
base numeral systems must represent information on geometrically spaced scales
differing by powers of say 2 or 10. The largest register must represent a scale
comparable to the largest representable number and the smallest register scale
must be comparable to the finest resolvable detail. The ability of an RNS to use
closely spaced registers to represent a large range of positions is critically
useful for position encoding in dMEC, with its narrowly distributed lattice
periods. 


Second, addition, subtraction, and multiplication are completely parallelized in
an RNS, because unlike base numeral systems, they require no ``carrying-over'' of
information from one register to the next: the sum of two numbers is the modulo
sum of their residues computed independently within each register. In computer
science, this property has long been appreciated and is used to perform fast,
parallelized computations in signal processing and RSA applications
\cite{RNSreview, RSAref,RNSapps}. The moving rat, using an RNS, can update its estimate
of position by independently incrementing the phases of each lattice, without
carrying phase winding information from one lattice to another. Indeed, in dMEC
each lattice phase is independently updated as a function of rat position,
without evidence of jumps in phase when a smaller lattice completes one winding.

\begin{figure}[h] 
\centerline{\epsfig{file=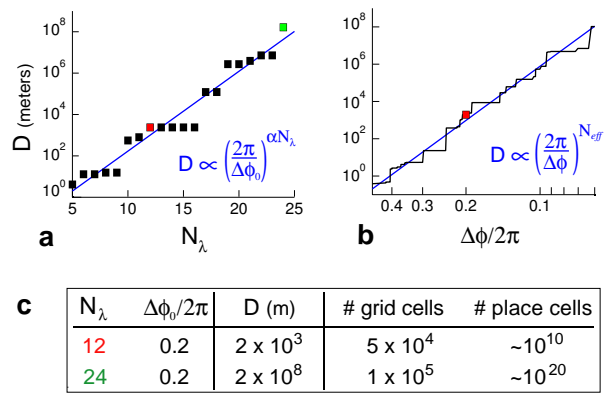,scale=0.5}} 
\caption[]{{\bf Capacity of dMEC under generalized RNS encoding.} The maximum
uniquely representable distance, $D$, grows exponentially as a function of
number of lattices, $N_\lambda$, {\bf a}, and as a power of the phase
uncertainty $\Delta \phi$, {\bf b}, even when the lattice periods are not co-prime and even if the phase uncertainty is equal across lattices. [The first lattice period is 30 cm, with 4 cm
increments per subsequent lattice. In {\bf a} ${\Delta \phi}/{2 \pi} = 0.2$, and
in {\bf b} $N_{\lambda} = 12$. The fit parameters are $\alpha \simeq 0.55$ and
$N_{\rm\emph{eff}} \simeq 9.7$. See Supplementary Information for more details.]
{\bf c} 12 lattices uniformly spaced from 30 cm to 74 cm with 5 resolvable
phases in each direction can unambiguously represent a $\sim$ 2 km $\times$ 2 km
area with 6 cm resolution. If $5000$ neurons build each lattice, that would
require $ \sim 5 \times 10^{4}$ neurons. To cover the area with sparse
unimodal place-cell like encoding (with 10 neurons per (6 cm)$^2$ block) would 
require $\sim 10^{10}$ neurons, compared to the estimated $10^5$ 
neurons in rat dMEC \cite{ECnumber1,ECnumber2}, which could sparsely represent 
at most $\sim$ 6 m$\times$6 m with (6 cm)$^2$ resolution. With 24 lattices, $D \sim 2 \times 10^5$ km
in each direction, hugely in excess of the representational requirements of
rats.}
\label{fig:scaling}
\end{figure}

Some important properties of dMEC encoding differ from a standard RNS. Lattice
periods in dMEC are probably not co-prime integers, but real numbers, as is rat
position. Also unlike an RNS, where the number of residues (distinguishable
phases or phase resolution) grows with modulus size (lattice period), the phase 
resolution in dMEC seems constant across lattices. Under these conditions, existing exact formulas for reconstructing numbers from their residues using the CRT no longer apply, and the representational capacity may be much smaller than possible in a standard RNS. But despite these
differences, summation and multiplication are still fully parallelized, and 
we show without recourse to any particular decoding scheme, that the 
representational capacity continues to scale
exponentially with the number of (non co-prime) lattices, spanning a vast range
of real-valued positions with high resolution (Figure 1 and Supplementary Information). The 
uniqueness of this representation guarantees that an inverse map from lattice
phases to position still exists. The capacity (Figure 1) under a generalized RNS encoding is  
far in excess of the navigational requirements of rats, and the surplus
capacity could be devoted to redundancy for error correction \cite{RNSreview,RNSapps} and robustness.

To summarize our argument, we suggest that dMEC treats position simply as a
2-coordinate vector, and then encodes the coordinates, two real \emph{numbers},
using a unique numeral system that can represent large numbers with a narrowly
spaced set of registers. At first glance, one might have imagined that dMEC firing 
reflects some kind of integral transform (e.g. Fourier or Wavelet) of 2-dimensional space 
and the rat's location in it. To do this, dMEC would have to treat rat position as an entire 2-d \emph{function}, such as a blob centered at the rat's location in space, and transform this into a set of 2-d patterns of different spatial frequency. Besides the computational unnecessity of representing two position coordinates by an entire function, the integral transform scheme is unlikely for at least two reasons:  (1)  It is incompatible with the data. The experiments \cite{GridCells1, GridCells2, Sharp99} unequivocally show that the patterns of dMEC activity are stereotyped, and independent of enclosure size or shape; given the activity pattern of one neuron in each lattice in any one enclosure, the activities of the rest are well specified upto a spatial phase shift in all enclosures.  Thus, dMEC encoding contains little additional information or flexibility to describe general functions in 2-d space, as required of integral transforms. (2) A Fourier-like transform cannot be used to represent position unambiguously over a distance greater than the largest lattice period ($\sim 2$ m). To reconstruct a single bump of rat position over a range $D$ with resolution $D_{min}$ would require lattice periods ranging from smaller than $D_{min}$  up to $D$. 

Our focus here has been on characterizing the general theoretical properties of
position {\em en}coding in dMEC. These insights provide the necessary foundation for understanding how information encoded in dMEC may be used by areas that receive dMEC inputs. 
We have illustrated, without regard to specific {\em de}coding schemes, how the maximal
capacity of dMEC for unambiguous position encoding scales as a function of dMEC phase
resolution and lattice numbers, and have observed based on the properties of an
RNS and our simulations, that these phases theoretically contain enough
information to uniquely encode position over a combinatorially large range. As we discuss next, our interpretation of dMEC encoding raises several
testable questions with specific implications for theory and experiment.

The most basic of these is whether the rat does actually make use of the range
of information we have shown is theoretically available in dMEC under an
RNS-like scheme. There are two (not mutually exclusive) ways in which rats may
use this information: (1) For homing and path integration over large ranges by
decoding the relative dMEC phases to compute rat displacements, or (2) For
attaching unique `labels', as described below, to a large number of specific
locations, with landmark-independent path integration only between nearby
locations. Either scenario can be probed experimentally, as we describe below, 
but in both cases resolving the fundamental issue requires testing in 
enclosures of size approaching the behaviorally relevant range for rats. 

In the first case, assuming dMEC is the primary source of position
representation, our proposal would be strongly supported if the rat
can perform reasonably accurate landmark-free homing behaviors over ranges 
much larger than the largest dMEC lattice period. This scenario would require an explicit 
decoding of the phase encoding of position.  Because position encoding by modulo residues, which we argue is happening in dMEC, is a 1-1 onto (bijective) function over the illustrated 
range (Figure 1) of positions, the mathematical inverse exists and is unique. However, there could be numerous possible neurobiological decoding schemes to exactly implement this inverse or approximate it.  Our capacity estimate, which contains the experimentally observed phase uncertainty, bounds how much information can be extracted from dMEC and should be used as a guide for evaluating the efficiency of different decoding
schemes.  Experimentally, 
the test of the long-range position encoding hypothesis involves measuring the largest lattice 
period in dMEC, and quantitatively determining the range over which rats can 
perform accurate homing in large featureless spaces. 

In the second case, the large set of unique dMEC phases as a function of rat
position may be used as absolute markers for specific landmarks or positions in
familiar environments. In this scenario, the rat may {\em not} explicitly decode
position or displacements over distances larger than a lattice period, but may
essentially use the vast set of distinct dMEC phases to uniquely represent a
large number of distinct, spatially separated locations. The rat may locally
perform path integration by updating phases, to ensure that starting from an
absolute phase at landmark A, the rat can take any path to landmark B and yet
obtain the correct absolute phase for B. Such a system would be useful to the
rat in distinguishing a familiar landmark from a lookalike but spatially
distinct location \cite{Kuipers00}. This proposal could be tested experimentally
by checking at multiple familiar landmarks in large, partially occluded
enclosures whether the absolute dMEC phases for each landmark are reproducible
across trials. If dMEC is reset to a set of previously assigned phases for landmark B upon reaching 
B from a novel occluded path and after extensive exploration in an unfamiliar landmark-rich environment, it would suggest that absolute phases are
important; the RNS scheme would allow dMEC to uniquely encode a large number of
landmarks.

Unlike most numeral systems, in an RNS scheme the representation of space by the different registers or dMEC lattices is not hierarchically ordered: all lattices are roughly
equal in their contribution to position representation at any scale. Therefore,
if hippocampal place cells are involved in reconstructing a position estimate or 
reading out a unique label of a landmark from dMEC, each local group of hippocampal cells 
must pool input from \emph{many} different dMEC lattices; indeed, neurons in the septal 
(dorsal) half of the long axis of the hippocampus do appear to receive inputs from across the 
dMEC band \cite{hippo_retrograde}. In addition, position reconstruction from
an RNS is susceptible to characteristic deficits following microlesions that successively 
destroy individual dMEC lattices. Error correction and redundancy could grant the system 
resilience against partial lesions, but beyond a critical
point, there should be a sudden degradation in position estimation or location
representation at all scales. By contrast, in any scheme where different
lattices encode positions at their corresponding spatial scales, selectively
lesioning the smallest (largest) lattices is likely to destroy fine (coarse), but not coarse (fine),
position estimation. 

When rats are trained to explore small enclosures or 1-d tracks, neurons in the hippocampus  
-- the primary output target of dMEC, and a critical locus of spatial learning and memory --
tend to form unimodal place-fields that resolve position with relatively high accuracy, and cover 
the space \cite{OKeefe78, Wilson}.  Based on our analysis, we predict that the 
inefficiency of sparse encoding (Figure 1c) and limited neuron numbers in 
hippocampus ($\sim 10^5-10^6$ neurons) \cite{ECnumber1}, rule out the 
possibility that hippocampus could fully remap large enclosures with high 
resolution using unimodal place-cells. Rather, driven by representational 
limitations in large areas, place cells must either generically develop multi-peaked responses as a 
function of rat position to cover space with reasonably high resolution, akin to dMEC neurons, or must disengage from finely representing all space with narrow place fields, instead covering only select locations based on other salient cues or associations. The latter scenario would make hippocampus a selective consumer and processor of position information from dMEC, leaving it free to perform more general associative tasks \cite{Markus95, Eichenbaum99, Sharp99}.

The possibility that dMEC may be representing position, a
continuous metric variable, using a compact and parallelized \emph{numeral
system} amenable to arithmetic operations such as addition or shifts on the
variable, is itself extraordinary. Such a numeral system code is qualitatively
different from all other known examples of coding in the brain \cite{codingreview}:
Proportional-rate coding (e.g., eye position and firing rate in the oculomotor
system \cite{LopezBarneo82}), unary or sparse coding (e.g., head direction cells \cite{Taube90}, place cells \cite{OKeefe78}), or even other kinds of combinatorial codes which 
represent non-metric variables (e.g.
odor \cite{Laurent01}) and therefore lack an arithmetic aspect. This encoding scheme 
provides insight into the ingenuity of neural codes, and provokes questions 
at the intersection of neuroscience, mathematics, and computer science.

{\bf Acknowledgments} We are grateful to Torkel Hafting, Michael Stryker, Loren Frank, Behrooz Parhami, David Tank, Sebastian Seung, and Uri Rokni for helpful discussions and comments. 

\end{document}


\title{Rat position encoding in triangular lattice neurons:  Supplementary Material}

\maketitle

We numerically estimate the range of positions that the phases of a small set of lattices, with spacings distributed over a narrow range, can unambiguously represent.  We assume that unlike a standard RNS, the dMEC lattice spacings are not co-prime integers, and the encoded position is a real number. There is no clear analytical statement for how the representational capacity of a generalized RNS, with limited, constant phase resolution across lattices and non co-prime lattice spacings, compares to that of a standard RNS, but we show without regard to any particular readout scheme that the capacity is large, and scales similarly to a standard RNS. For simplicity and clarity of exposition, most computations are in one dimension, but can be extended to two dimensions. 

{\bf Exact numerical evaluation} Consider a set of lattice spacings or modulii, 
$\{ \lambda_i | i=1,...,N_\lambda\}$. A position $x_1$ is represented by the vector of its residues, 
\[
x_{1i} = x_1 \ {\rm mod} \ \lambda_i, ~~~~\ i=1,...,N_\lambda,
\]
or phase vector $\phi_1$, whose entries are 
\[
\phi_{1i} = \biggl(\frac{2 \pi x_{1i}}{\lambda_i}\biggr) \ {\rm mod} \ 2 \pi.
\]
For example, when $x_1 \rightarrow x_1+\lambda_k$, then $\phi_{1k} \rightarrow (\phi_{1k}+2 \pi) \ {\rm mod} \ 2\pi = \phi_{1k}$. Assume that each phase can only be resolved upto a spread $\Delta \phi$. Then the lattice phases can be used to distinguish between $x_1$ and $x_2$ if there is a mismatch of more than $\Delta \phi$ in at least one entry of their phase vectors. 

Therefore, we define the phase distance between $x_1$ and $x_2$ to be the largest phase mismatch across lattices (entries) in their phase vectors: 

\begin{eqnarray}
\label{eq:distance}
\parallel {\bf \phi}_1-{\bf \phi}_2 \parallel = {\rm max_i}\bigl[{\rm min} \bigl[ &(\phi_{1i}-\phi_{2i}) \ {\rm mod} \ 2 \pi, \\ 
&(2\pi - \phi_{1i}+\phi_{2i}) \ {\rm mod} \ 2 \pi\bigr] \bigr]
\nonumber
\end{eqnarray} 

In the numerics for our figure, the smallest lattice spacing is 30 cm, and each subsequent 
lattice spacing is 4 cm greater than the last.  We conservatively assume a phase resolution 
of 1/5 of a period, or $\Delta \phi = 2\pi/5.$ We start with $\phi_i = 0$ for $x=0$, and 
increase $x$ in small increments to generate Figures 1a and 1b. The range $D$ in Figure 1a 
is defined as the point at which the phase distance between $x=0$ and $x=D$ drops to within 
the phase resolution, $\Delta \phi$. 
Figure 1b plots the phase resolution needed (or maximum 
allowed phase uncertainty) to cover the range $x=0$ to $x=D$ as a function of $D$, for a 
fixed number of lattices ($N_\lambda=12$). 

{\bf Scaling ansatz for generalized RNS} 
In a standard RNS, the range of sequential numbers that can be 
distinguished from each other is bounded by the number of 
distinct states of the $N_{\lambda}$ registers,
\begin{equation}
N_{\rm max} = \lambda_1\cdot \lambda_2 \cdots \lambda_{N_{\lambda}}.
\label{eq:Nmax}
\end{equation}
The representable range reaches this upper bound if 
the registers are co-prime integers (Otherwise the representable range is the least common multiple of the registers). To speak of
distances instead of integer numbers, we may associate a length scale
of unity to the finest distance resolution, $D_{\rm min}=1$. The  
maximum representable range is then $D_{\rm max}= N_{\rm max}$, as given
by Eq.~(\ref{eq:Nmax}).
If all the lattice periods $\lambda_i$ are similar in size, then the maximum range scales 
exponentially with $N_\lambda$ and as a power of the distinct number of states in each 
register ($\sim \lambda$): 
\[
\nonumber
N_{\rm max} \sim \lambda^{N_\lambda}. 
\]

In the more general case where the $\lambda_i$ are real numbers
with a finite resolvable phase resolution, a similar
counting argument can be made as follows. Assume that the
smallest measurable increment in phase
$\Delta \phi$ is the same for all lattices, and
divide the
phase interval, $0$ to $2\pi$, into
bins of size $\Delta \phi$. For the purpose of our scaling estimate (and unlike in the 
exact numerical calculation described above), phase is thus 
discretized,
into $2 \pi/\Delta \phi$ bins which correspond 
to the approximate
number of resolvable states of each lattice
\footnote{Note that the distance measure in Eq.~(\ref{eq:distance}), used for the exact
numerical calculation, is written in terms of continuous phases and unlike in the scaling estimate here does not involve any binning.}. 
The total number of states of all lattices is
\[
N_{\rm max} = \left( \frac{2 \pi}{\Delta \phi}\right) ^{N_\lambda}
\]

The finest resolvable change in position is 
$D_{\rm min} = \lambda_1 \Delta \phi/2\pi$, where $\lambda_1$ is 
the smallest lattice spacing. As the position is incremented
in steps of $D_{\rm min}$, the phases of all lattices necessarily
go back to their original state within $N_{\rm max}$ steps. Hence
the maximal representable range is bounded by 
\begin{equation}
D_{\rm max} = \lambda_1 \left( \frac{2 \pi}{\Delta \phi}\right)^{N_\lambda-1}
\label{eq:counting}
\end{equation}
\noindent As in standard RNS, the maximal combinatorial capacity may 
be achieved only for particular choices of $\lambda_i$. For generic choices of the lattice periods $\lambda_i$, the representable range
may not reach the bound of Eq.~(\ref{eq:counting}); nevertheless, we still expect the  range $D_{\rm max}$ to scale exponentially with the number of lattices, and algebraically
with the number of states in each lattice: 
\begin{equation}
D_{\rm max} \propto \left(\frac{2\pi}{\Delta \phi}\right)^{\alpha N_{\rm lambda}}
\label{eq:ansatz}
\end{equation}
where the parameter $\alpha$ is expected to be of order unity. 
This relation is an ansatz, which we test by comparison with 
the numerical results for particular sets of the lattice spacings.
We find that the maximum range typically does scale according to the 
ansatz, as demonstrated Fig.~1 (solid lines in parts a and b).
In part a, $D_{\rm max}$ scales roughly exponentially with $N_\lambda$,
and $\alpha \simeq 0.55$. In part b the
dependence on phase resolution is shown
for a particular number of lattices, $N_\lambda = 12$. In
agreement with the ansatz, 
$D_{\rm max}$ scales roughly as a power of $2\pi/\Delta\phi$,
with an exponent $N_{\rm eff} \simeq 9.7$.

In summary, our numerical results imply that capacity scales algebraically with phase resolution and
exponentially with the number of lattices.  Interestingly, this implies that to achieve a large 
capacity with a fixed number of neurons, neurons should be devoted to building more lattices rather 
than increasing the phase resolution: this result may help to explain the surprisingly poor phase
resolution found in dMEC, where neural activity blobs cover 1/3 of the total lattice period. 
